\documentclass[a4paper,12pt]{article}
\usepackage{graphicx}

\usepackage{amsmath,amssymb,graphicx}
\usepackage{cite}
\usepackage{color}
\usepackage{color}
\usepackage{ifpdf}
\ifpdf %if using pdfLaTeX in PDF mode
  \DeclareGraphicsExtensions{.pdf,.png,.jpg,.jpeg,.mps}
  \usepackage{pgf}
  \usepackage{tikz}
\else %if using LaTeX or pdfLaTeX in DVI mode
  \usepackage{graphicx}
  \DeclareGraphicsExtensions{.eps,.bmp}
  \usepackage{pgf}
  \usepackage{tikz}
  \usepackage{pstricks}
\fi
\usepackage{epic,bez123}
\usepackage{floatflt}% package for floatingfigure environment
\usepackage{wrapfig}% package for wrapfigure environment
\numberwithin{equation}{section}

\newcommand{\dd}{\mathrm{d}}

\begin{document}

\title{Action-angle variables in curved space-time}%
\author{Jaehun Lee\thanks{e-mail:\texttt{ljhiverson@gmail.com}}, and 
Corneliu Sochichiu\thanks{e-mail:
\texttt{corneliu@gist.ac.kr}}\\
{\it GIST College, Gwangju Institute Science and Technology} \\
{\it 123 Cheomdan-gwagiro, Buk-gu, Gwangju} \\
{\it Republic of KOREA}
}%
%
%\subjclass{}%
%\keywords{}%

%\date{}%
%\dedicatory{}%
%\commby{}%
% ----------------------------------------------------------------
\maketitle
\begin{abstract}%:abstract
 We construct a relativistic and curved space version of action-angle variables for a particle trapped in a gravity and electromagnetic background with time-like isometry. As an example, we consider a particle in AdS background. Furthermore, we obtain the semiclassical quantisation of its energy levels.
\end{abstract}
%\maketitle
%\tableofcontents
% ----------------------------------------------------------------
\section{Introduction}\label{sec:Intro}%:Introduction
% --------------------------------------------------------------------
Action-angle variables give a parameterisation of a classical system with finite motion in which one variable (action) remains constant, while its canonical conjugate (angle) is evolving linearly in time. 

At the early stage of quantum mechanics the discrete nature of energy was explained by the property of action variable to take only discrete values, an approach known today as Born-Sommerfeld quantisation  \cite{sommerfeld1921atombau}. So, the action-angle variables were studied extensively and became a standard subject for major classical mechanics textbooks  (see e.g. \cite{goldstein2014classical}). We now know that although in some cases the Born-Sommerfeld prescription gives exact energy eigenvalues, it is generally not a proper way to describe a quantum system. Even so, Born-Sommerfeld quantisation remains a valid approximation when quantum effects are not very strong. Hence, action-angle variables are a powerful and intuitive tool for semiclassical analysis.

Beyond the application to semiclassical analysis the set of action-angle variables is an interesting construction by itself because it explicitly reveals the structure of a classically integrable system with finite motion. As the approach was mainly developed in the era of non-relativistic physics, application to relativistic systems was largely overlooked.

Here we address the question of whether a relativistic (re)formultaion of the action-angle variables is also possible?  

Although, there is a number of applications of action-angle variables to relativistic particles in gravity backgrounds
(see \cite{Galajinsky:2013osa,Galajinsky:2013mla,Saghatelian:2014uba} for recent works), the method is applied in situations when the original relativistic system is equivalently reformulated in a non-relativistic manner.
So, in spite of very basic nature of this original question, we did not find a satisfactory answer in the literature.

Here we consider a relativistic particle moving in a curved \( (1+1) \) - dimensional space-time background with a time-like Killing vector. In addition, the particle is charged and interacts with a background electromagnetic field with Killing vector invariant strength. We explicitly construct (in quadrature) the action-angle variables and give the (implicit) formula for semiclassical Born-Sommerfeld quantisation of energy levels for the particle trapped by the background. 

The organization of the paper is as follows. In the next section, we introduce our approach starting from the discussion of geometric criteria of applicability of our formalism and further deriving the quadrature formulas for action and angle variables. There we also give the semi-classical quantisation prescription. Then we consider the example of two-dimensional Anti-de Sitter (AdS) space (or radial part of a higher-dimensional AdS space) and obtain quantised semiclassical energy levels for a particle trapped in such a space. Finally, we discuss our results.

% --------------------------------------------------------------------
\section{Relativistic particle in a background with time-like Killing vector}\label{sec:sec}%:Section
The charged particle in a curved space-time is described by the following action,
\begin{equation} 
	\mathcal{S}=\int \{-m\sqrt{g_{\mu\nu}\dot{x}^{\mu}\dot{x}^{\nu}}-e\dot{x}^{\mu}A_{\mu}\}\dd \tau\;,
\end{equation}
where $g_{\mu\nu}$, is the \( 1+1 \)-dimensional metric, and \( A_{ \mu} \), respectively, the vector potential of two-dimensional Abelian gauge field. Greek indices \( \mu, \nu \), etc. run through the range \( (0,1) \).

In classical mechanics action-angle variable method is applied to one--dimensional particles moving in static binding potential. The notion of `binding' we will clarify later, but the geometric analogue of static potential is existence of an isometry given by a time-like Killing vector field. In our situation we also require the Killing vector to commute with the gauge field strength.

The time-like isometry together with diffeomorphism invariance allow us to choose a coordinate system in which metric is time-independent and diagonal, i.e. the only non-trivial components of metric in  \( (t,x) \)-coordinates are \( g_{00}(x) \) and \( -g_{11}(x) \).\footnote{Notice, that we separate the sign from the definition of \( g_{11} \), i.e. \( g_{11}(x)>0 \).}

As for the gauge potential \( A_{ \mu} \), we can impose the gauge condition, \( A_{1}=0 \). In our coordinate system the remaining component \( A_{0} \equiv \phi(x) \) will be time-independent as long as the field strength is time-independent.\footnote{The proof is left to the reader.} The last is precisely the required above Killing invariance of the field strength tensor.

Fixing the world-line time reparametrisation by choosing,
\begin{equation}
 	\tau = t(\equiv x^{0}),
\end{equation}
we bring the action to the form,
\begin{equation}\label{Action:stand}
\mathcal{S}=\int \{- m\sqrt{g_{00}-g_{11}\dot{x}^{2}}-e\phi(x)\}\dd t .
\end{equation}
The action \eqref{Action:stand} is the starting point of our analysis. 

Before starting the analysis, let us observe that with a space-like coordinate transformation \( x'=x'(x) \), where
\begin{equation}
 	x'(x)= \int \sqrt{ \frac{g_{11}}{ \sqrt{ g_{00}}}} \dd x,
\end{equation}
we can bring the metric to the form \( g_{00}=g_{11}^{2} \). In such a coordinate system, the non-relativistic limit of the action is particularly natural and easy,
\begin{equation}\label{nonrel-act}
 	\mathcal{S}_{\text{non-rel}}=
	\int \left(  \frac{1}{2}m \dot{ x'}^{2}- m\sqrt{g_{00}}-e \phi \right) \dd t + \dots,
\end{equation}
where \( g_{00}= g_{00}(x(x')) \) and \( \phi= \phi (x(x')) \).
Hence, in this parameterisation our system is approaching a standard non-relativistic massive particle with the potential,
\begin{equation}\label{nonrel-pot}
 	V(x')= m\sqrt{ g_{00}}+e \phi,
\end{equation}
as  long as  \( \dot{ x}'{}^{2}\ll \sqrt{g_{00}} \). 
Dots in eq. \eqref{nonrel-act} denote terms which are small in this limit. 
% --------------------------------------------------------------------

\section{Action-angle variables}

Let's start by the Legendre transform of the action \eqref{Action:stand} to the Hamiltonian description. The canonical momentum is given by,
\begin{equation}\label{Leg:momentum}
 	p \equiv \frac{ \partial \mathcal{L}}{ \partial \dot{x}}=
	\frac{m g_{11}  \dot{ x}}{ \sqrt{g_{00}- g_{11}\dot{ x}^{2}}}.
\end{equation}

The expression \eqref{Leg:momentum} can be easily reversed  for the velocity,
\begin{equation}
 	\dot{ x}= \sqrt{ \frac{g_{00}}{g_{11}}}
	\frac{p}{ \sqrt{p^{2}+m^{2}g_{11} }}.
\end{equation}

The Hamiltonian is then given by,
\begin{equation}\label{Hamiltonian}
	H \equiv p \dot{x}-\mathcal{L}=  \sqrt{ \frac{g_{00}}{g_{11}}}
	\sqrt{ p^{2}+ m^{2} g_{11}}
	+ e \phi. 
\end{equation}

Now let us apply the Hamilton-Jacobi method to solve this system. Recall that the idea of the method is to make a canonical transformation to new variables \( J \) and \( \theta \), such that the new canonical momentum \( J \) is cyclic. The characteristic function \( W(x, E) \) of the transformation is found from the condition, 
\begin{equation}\label{HJeq}
	H(x,\partial W/ \partial x)-E =0.
\end{equation}
where $E $ is a constant depending on initial conditions and determining the value of the Hamiltonian. The formal solution to the eq. \eqref{HJeq} is given by,
\begin{equation}\label{Char-func}
 	W=
	\int_{x_{0}}^{x} 
	\sqrt{ \frac{g_{11}}{g_{00}}(E - e \phi)^{2}-m^{2} g_{11}}\,\dd x.
\end{equation}
where \( x_{0} \) is the initial value of position.

In the case of compact motion, we can define the \emph{action variable} by extending the integral defining the characteristic function \eqref{Char-func} to one periodicity cycle of motion,
\begin{equation}\label{action-var}
 	J \equiv\oint p \dd x=\oint \sqrt{ \frac{g_{11}}{g_{00}}( E- e \phi)^{2}-m^{2} g_{11}}\,\dd x.
\end{equation}

The integration path in \eqref{action-var} is bounded by \emph{classical turning points}, at which the integrand vanishes, i.e., 
\begin{equation}\label{eq:tp}
 	E-e \phi(x)= \pm m\sqrt{g_{00}}.
\end{equation}
The existence of turning points determines the bound motion. As the particle at turning points is in a non-relativistic regime, the turning point condition, curiously, is  exactly the same as for a non-relativistic system with potential \eqref{nonrel-pot}, apart from the possibility of sign variation.

The turning points \( x_{i} \) and \( x_{i+1} \) are  bounding a classical region determined by the condition,
\begin{equation}
 	 m\sqrt{g_{00}} \leq |E- e \phi|.
\end{equation}

Notice, that for non-vanishing mass the two branches, corresponding to either choice of sign in \eqref{eq:tp} are well separated. This implies that both ends of the same classical region should have turning points corresponding to the same choice of the sign, which means that a particle can never ``turn back in time'' within a classical region. This is a manifestation of particle/antiparticle conservation in relativistic mechanics. In the case of zero mass, however, the situation could be different.\footnote{In this case one should consider non-charged particles.}  

In a generic reference frame with the Killing vector given by components \( \xi^{ \mu} \), where \( \xi^{2} =\xi^{ \mu} \xi_{ \mu}>0 \) the general covariant form of the action variable \eqref{action-var} is given by,
\begin{equation}\label{action-var-gi}
 	J(E)=
	\oint_{C_{ E,\xi}} \sqrt{(E-e \xi \cdot A)^{2}/ \xi^{2}-m^{2}} \dd \lambda,
\end{equation}
where \( \dd \lambda \) is the invariant integration measure over the `cycle' \( C_{E, \xi} \) given by  \( x^{ \mu}( \tau) \) such that,
\begin{equation}
 	\xi_{ \mu} \dot{ x}^{ \mu}=0,
\end{equation} 
and condition,
\begin{equation}
 	E-e \xi^{ \mu} A_{ \mu} \geq m.
\end{equation}
The expression \eqref{action-var-gi} is manifestly reparametrisation invariant, while  in the special coordinate frame in which the Killing vector is \( \xi^{ \mu}=(1,0) \) we recover the eq. \eqref{action-var}.\footnote{In this coordinates the covariant Killing vector is \( \xi_{ \mu}= (g_{00},0) \).} Notice, that \( A_{ \mu} \) is still gauge fixed.

Classical equations of motion imply that the action variable \( J \) take constant values along the classical path. Let us solve the Eq. \eqref{action-var} for \( E \).  Then, the conjugate angle variable \( \theta \) satisfies the equation,
\begin{equation}
 	\dot{ \theta}= \frac{ \partial E (J)}{ \partial J} \equiv \omega (J).
\end{equation}
As \( \omega(J) \) is a constant of motion too, the solution for \( \theta(t) \) is given by,
\begin{equation}
 	\theta(t)= \omega t+ \theta_{0},
\end{equation}
where \( \theta_{0} \) is the (new) initial condition, which together with the value of \( J \) (or energy \( E \)) gives a complete set of initial conditions.

% --------------------------------------------------------------------
%\subsection{Bohr-Sommerfeld quantization}

Now we are ready to consider the semiclassical quantisation of the system. According to Bohr-Sommerfeld quantisation prescription the action variable should take discrete values given by,
\begin{equation}
 	J(E_{n})= 2\pi (n+ \gamma)\hbar,
\end{equation}
where \( n=0,1,2\dots \), and  \( 0\leq \gamma <1 \) is a constant determining the vacuum value of the action variable. This gives an implicit formula for the energy level \( E_{n} \). 
% --------------------------------------------------------------------

% --------------------------------------------------------------------
\subsection{Example: AdS space}
 There are many good examples to play with our approach, among which the Anti-de Sitter space (AdS) is, perhaps, distinguished due to importance of this geometry.
 
The metric of the \( D+1 \)-dimensional AdS is given by,
\begin{equation}\label{AdS-metr}
	\dd s^{2}=\left(1+\frac{r^{2}}{R^{2}}\right)\dd t^{2}-\left(1+\frac{r^{2}}{R^{2}}\right)^{-1}dr^{2}-r^{2}\dd\Omega_{D-1}^{2},
\end{equation}
where \( \dd \Omega_{D-1} \) is the differential of \( (D-1) \)-solid angle (the hyper-volume of a unit \( (D-1) \)-sphere).

In the case of \( (1+1) \)-dimensional space or a pure radial motion, the angular part can be discarded. In this case we deal with a two-dimensional space with the metric,
\begin{equation}
	\dd s^{2}=\left(1+\frac{r^{2}}{R^{2}}\right)\dd t^{2}-\left(1+\frac{r^{2}}{R^{2}}\right)^{-1}\dd r^{2},
\end{equation}
which is in the ``standard'' form of Eq. \eqref{action-var}.

Applying directly the definition \eqref{action-var}, we find the action variable,
\begin{equation}\label{JAdS}
 	J=
	\oint \sqrt{ \frac{E^{2}}{(1+r^{2}/R^{2})^{2}}- \frac{m^{2}}{1+ r^{2}/R^{2}}} \dd r.
\end{equation}
The integration is over one cycle of motion bounded by turning points \( r_{0} \) determined by the equation,
\begin{equation}
 	E^{2}-m^{2}\left( 1+ \frac{r^{2}}{R^{2}}\right)=0 \Rightarrow
	r_{0} = R \sqrt{(E/m)^{2}-1},
\end{equation}
i.e. a particle in the AdS space is trapped in the region \( r \leq R \sqrt{(E/m)^{2}-1} \). Obviously, its energy can not be less than its mass\dots

Evaluation of the integral in Eq. \eqref{JAdS} yields,
\begin{equation}
 	J=2\pi R(|E|-m).
\end{equation}
Solving this for the energy we get,
\begin{equation}
 	|E|=m+J/2\pi R.
\end{equation}
This gives the angular speed \( \omega_{\text{AdS}}=1/R \), and complete classical description of the model. 

The semiclassical energy levels are given by,
\begin{equation}
 	|E_{n}|=m+ \frac{ \hbar}{R} (n+ \gamma),
\end{equation}
which are pretty much compatible with the result of solving the Schr\"{o}dinger equation in AdS, \cite{Fitzpatrick:2014vua} (see also \cite{Kapl}). Let us note, that positive energies (\( E_{n}>m \)) correspond to energy levels of particle, while negative energies (\( E_{n}<-m \)) stand for the anti-particle.

Remarkably, the particle in the AdS space background provides an example, alongside the harmonic oscillator, for which semiclassical energy levels are exact.

% --------------------------------------------------------------------

\section{Discussion}

In this work, we introduced the action-angle variables for a relativistic particle moving in gravity and electromagnetic field background. The approach is readily available as long as background is static, i.e. there is a time-like Killing vector field commuting with the metric and the electromagnetic field strength. The existence of time isometry is the relativistic counterpart of the conservativeness of a non-relativistic system. 

 As an example, we apply the approach to the radial motion of a particle in the anti-de Sitter background for which we were able to find semi-classically quantised energy levels. Let us note that when \( E\gg m \), the classical trajectory of the particle is deeply relativistic. Therefore the non-relativistic approximation couldn't be used here, while the system can be still semiclassical.
 
 Although the method is explicitly constructed for 1+1-dimensional spaces, it can be generalised to higher dimensions as long as geometry allows separation of dynamical variables. 

% --------------------------------------------------------------------
\subsubsection*{Acknowledgements}
This work is  done within the undergraduate research program for the bachelor's degree thesis \cite{Jaehun-thes}.

% --------------------------------------------------------------------
\bibliographystyle{unsrt}
\bibliography{AAvariable}%:select bib file
	
\end{document}